# Analysis of a Sputtered Si Surface for Ar Sputter Gas Supply Purity Monitoring

Uwe Scheithauer, 82008 Unterhaching, Germany
Phone: + 49 89 636 – 44143 , E-Mail: scht.uhg@googlemail.com

**Abstract:**

For sputter depth profiling often sample erosion by $Ar^+$ ions is used. Only a high purity of the sputter gas and a low contamination level of the ion gun avoid misleading depth profile measurements results. Here a new measurement procedure is presented, which monitors these parameters. A Si sample is sputtered inside the instrument and then the surface concentration of the elements Ar, C, N and O is measured. Results of such measurements of an XPS microprobe PHI Quantum 2000, which were recorded over a period of 10 years, are presented.

______________________________________________

## 1. Introduction

For composition analysis of thin film systems often in-depth profiling is the method of choice. Applying in-depth profiling first the sample surface is eroded by ion bombardment ("sputtering") usually using inert gas ions with a kinetic energy between 250 eV and 5 keV. Then the residual surface is analyzed after each sputter erosion step utilizing Auger electron spectroscopy and X-ray photoelectron spectroscopy (XPS), for instance. The depth distributions of the elements are recorded as a function of sputter time [1].

The ISO committee TC201/SC 4 considers many aspects of sputter depth profiling [2]. In especially, aspects of ion beam alignment, the measurement of the ion beam intensity, the calibration of depth scales and the choice of adequate reference samples are discussed. But some more instrumental aspects of sputter depth profiling were not taken into account by the actual ISO standards. A new procedure, which monitors the vacuum quality of a sputter depth profiling instrument by the recontamination rate measurement of a sputtered Ti sample, was published recently [3]. Here a novel method to monitor the Ar sputter gas purity and to some extend the ion guns contamination level is introduced. Occasionally, for a monitor measurement a Si sample is sputtered by $Ar^+$ ions and the surface concentration of the elements Ar, C, N and O is used to characterize the instruments performance. Data of an XPS microprobe PHI Quantum 2000, which were recorded over a period of 10 years, are shown.

## 2. Instrumentation

For the measurements presented here a Physical Electronics XPS Quantum 2000 was used. This XPS microprobe achieves its spatial resolution by the combination of a fine-focused electron beam generating the X-rays on a water cooled Al anode and an elliptical mirror quartz monochromator, which monochromatizes and refocuses the X-rays to the sample surface. Details of the instruments design and performance are discussed elsewhere [3-10].

For sputter depth profiling the instrument is equipped with a differentially pumped $Ar^+$ ion gun. Sputter ion energies between 250 eV and 5 keV are selectable. Thermally grown $SiO_2$ on a Si wafer, whose thickness was estimated by ellipsometry, is used as reference material for sputter erosion rate calibration [11]. The Ar gas supply of the sputter ion gun is rather complex [12]: The Ar gas for the sputter ion gun is supplied in a 100 bar gas bottle. A pressure regulator reduced the Ar gas pressure. By a pneumatic valve, which is operated in pulsed mode, and a fixed metal leak the Ar gas is inject into a gas reservoir. The gas pressure in this reservoir is measured and regulated. Therefore either additional Ar gas is injected or the reservoir is evacuated from time to time with the turbo molecular pump utilizing a second pneumatic valve. The Ar gas is delivered from the reservoir to the ion gun by a small tube. All devices are connected by vacuum coupling radiation (VCR) fittings.

For flat mounted samples as used here in a Quantum 2000 the incoming X-rays are parallel to the surface normal. In this geometrical situation, the mean geometrical energy analyzer take off axis and the differentially pumped $Ar^+$ ion gun are oriented ~ 45° relative to the sample surface normal.

The samples were mechanically mounted on a 75mm x 75mm sample holder. This sample holder is introduced into the XPS vacuum chamber via a turbo pumped intro chamber.

## 3. Data Evaluation

Data evaluation was done by an improved version of the PHI software MultiPak 6.1 [13]. In case of measured peak intensities quantification it uses the simplifying model, that all detected elements are distributed homogeneously within the analyzed volume. This volume is defined by the analysis area and the information depth of an XPS measurement, which is derived from the mean free path of electrons [14]. Using this approach one monolayer on top of a sample quantifies to ~ 10 ... 30 at% depended on the samples details.

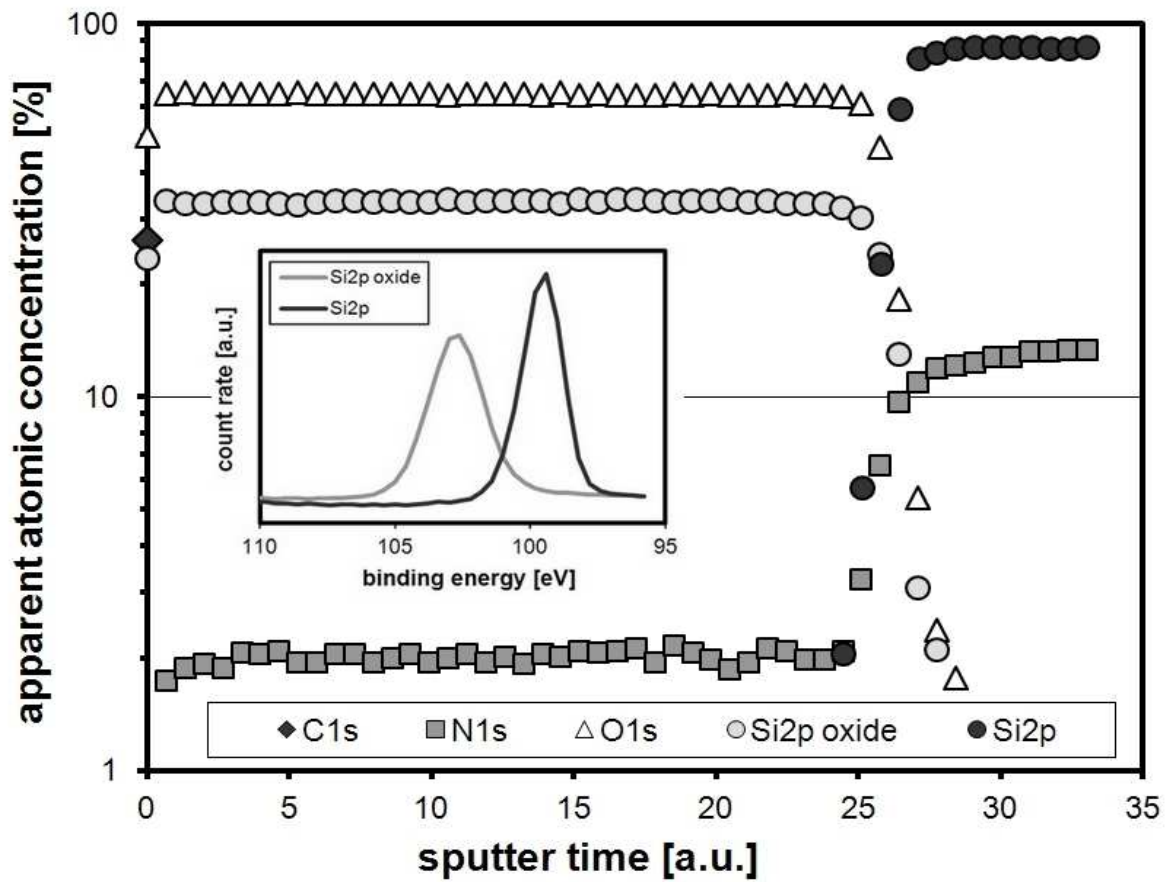

Fig. 1:   XPS depth profile measurement, 104.6 nm $SiO_2$ on Si substrate; date: 28.8.2003
Due to a leakage in the ion gun gas supply N was implanted.
For data evaluation nlls fitting was used. The insert displays the spectra of Si and SiO2, respectively.

The measured depth profile data were evaluated using non-linear least square (= nlls) fitting. Applying this evaluation method the data are fitted with internal reference spectra allowing small peak energy shifts. Each spectrum in each sputter depth is fitted by a linear combination of these reference spectra. If in the measured depth profile data the XPS spectra of one elemental signal are significantly shifted or are differently shaped, which represent different chemical surroundings of this element, the corresponding chemical depth distributions of this element are evaluated.

## 4. Experimental Results

Fig. 1 shows a depth profile measurement of a 104.6 nm $SiO_2$ layer on a Si substrate. For sputter depth profiling 2 keV $Ar^+$ ions were used. The $SiO_2$ was thermally grown on a Si wafer. The layer thickness was evaluated by ellipsometry. Measurements of this reference standard are utilized to calibrate the sputter rate of the ion gun from time to time. This depth profile shows unexpected results. In the $SiO_2$ layer ~ 2 at% N and in the Si substrate ~ 13 at% N were detected. The detection of implanted N gives a strong hit to an air leakage of the Ar gas supply. In this case a replacement of the pressure regulator of the 100 bar Ar bottle fixes the problem.

As described above, the Ar gas supply is constructed rather complex. A new measurement procedure, which can monitor the gas supply condition, is developed from the observation of the N implantation into the Si during the sputter depth profiling measurement discussed above (fig. 1). The absence of air leakages and the purity of the Ar gas are monitored by XPS measurements of a sputtered Si sample. Fig. 2 shows an example of such a measurement. For sputter depth profiling 2 keV $Ar^+$ ions were used. The O1s, N1s, C1s, Ar2p and Si2p signals were recorded immediately after sputtering. The inserted table summarizes the surface composition. Additionally the measurement time, which was used to record the XPS signals, is specified. Most of the time was spend to record the C1s, N1s and O1s peak. To obtain high count rates and thus a good detection limit, the data were measured utilizing a high power X-ray beam (~ 200 µm diameter, ~ 45 W electron beam power) and a low energy resolution of the energy analyzer. The N1s XPS peak intensity monitor air leakages of the Ar gas supply system. Additionally the C1s and O1s peak were measured. Both peak intensities give an insight to the contamination level of the sputter ion gun. But in cases, when many new mechanical components were mounted inside the UHV, these contamination also monitor the sample surface recontamination by adsorption from the UHV system [3].

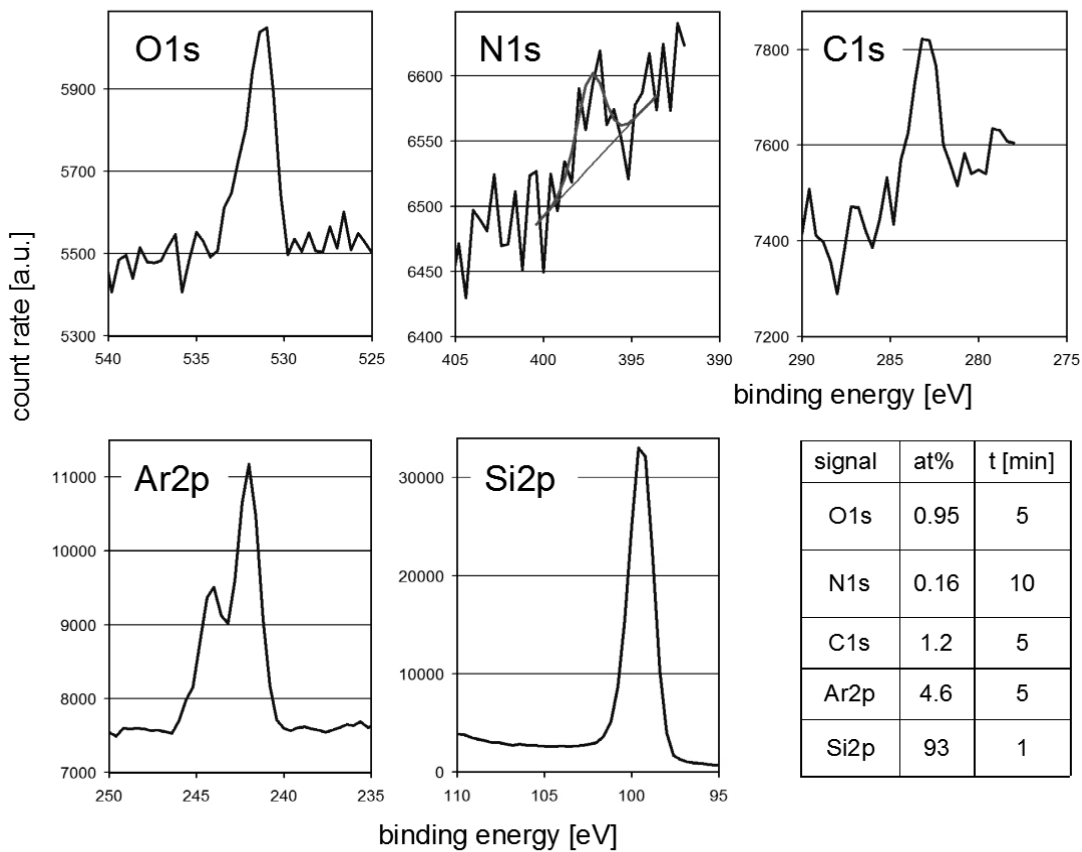

Fig. 2: example of the data recorded for regular monitoring of the ion guns condition
date: 21.1.2008 (after system bake-out / new ion gun filament)
The surface composition and the time used to measure each signal are given in the table.

Fig. 3 shows the results of sputter gas monitor measurements taken over a time period of 10 years. Against the measurement date the detected elemental concentration of Ar2p, C1s, N1s and O1s are plotted on a logarithmic scale. If an element was not detected, the detection limit is given by a hatched column. The detection limits vary a little bit, because the measurement procedure was modified during the first years and because of uncertainties, which are unavoidable if small signals are evaluated. The results of the depth profile (fig. 1) are shown (date: 28.08.2003). Only this time a serious leakage of the Ar gas supply was observed. The data of fig. 2 (date: 21.01.2008) were measured after the replacement of the ion gun filament and a bake-out of the whole system. Approximately one month later (date: 29.02.2008) the contamination level is reduced. The XPS microprobe was operated under UHV during this time. Since the replaced the new ion gun filament was heated all the time. It has outgased its contaminations during the one month period. These examples show how the data correlate with the system conditions and system modifications, respectively.

## 5. Conclusions

This article presents a new procedure, which allows monitoring the sputter gas purity and the condition of the $Ar^+$ ion gun. A Si sample is sputtered inside the instrument and then the surface concentration of the elements Ar, C, N and O is measured. Repeating the monitor measurements over a long time gives a reliable overview over the long time system conditions. The Ar, which was implanted into the Si, indicates a properly gas supply of the ion gun. Leakages of the Ar gas supply against air could be detected by an increase of the N signal. But also other aspects of the instruments history are reported by the data: The O and C concentration can monitor the contamination level of the ion gun and the whole instrument.

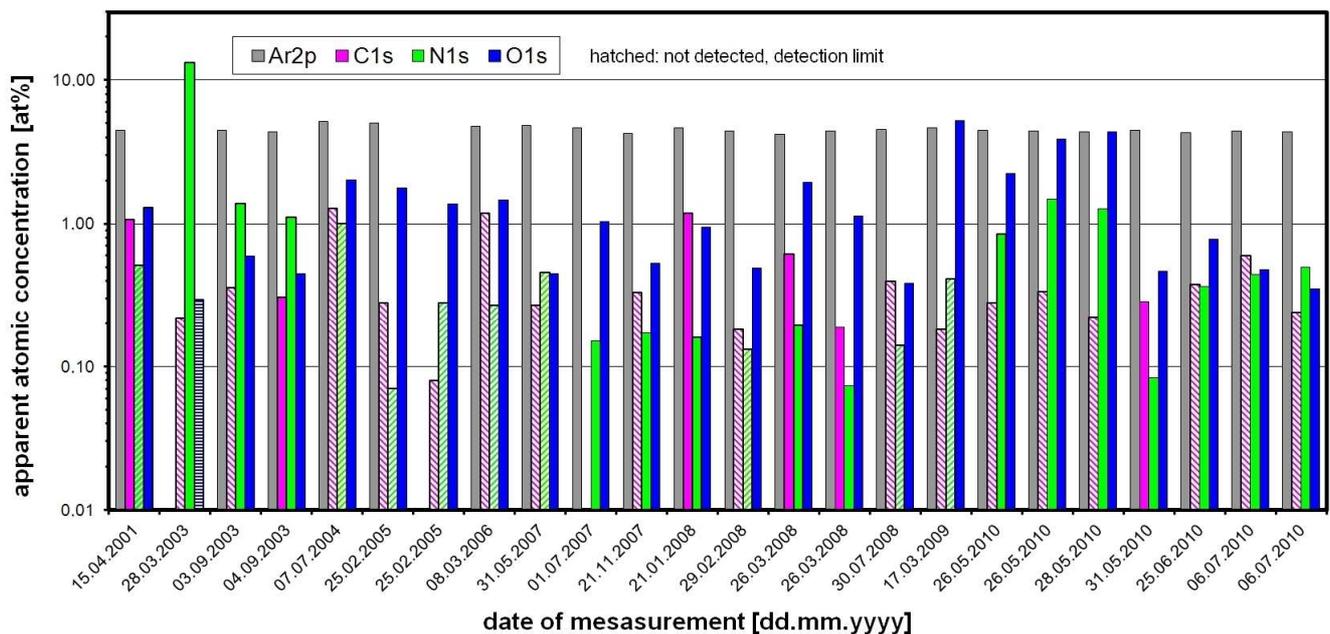

Fig. 3: results of long time monitoring

The procedure can easily be adapted to a sputter depth profiling instrument using an inert gas ion sputter gun. Obviously the necessary hardware for the measurement exists. Only a piece of Si is needed for the measurement. In a certified laboratory these measurements can be one module of the analysis quality management system.

## Acknowledgement

All measurements were done utilizing the Quantum 2000, instrument no. 78, installed at Siemens AG, Munich, Germany. I acknowledge the permission of the Siemens AG to use the measurement results here. For fruitful discussions and suggestions I would like to express my thanks to my colleagues.